\documentclass{article}  
\usepackage[dvips]{graphicx}
\begin{document}

\title{$\Theta^+$ pentaquark in the nuclear medium}

\author{M.~J. Vicente Vacas, D. Cabrera, Q.~B. Li, V.~K. Magas and E. Oset}
\maketitle

\begin{center}
\begin{small}
\it{ Departamento de F\'{\i}sica Te\'orica and IFIC,\\ 
Centro Mixto  Universidad de Valencia - CSIC, \\
Institutos de Investigaci\'on de Paterna, \\ 
 Apdo. correos 22085, 46071, Valencia, Spain}
\end{small}
\end{center}

\begin{abstract}{We study the interaction of the $\Theta^+$ pentaquark with nuclear matter
associated to the $KN$ decay channels and to the two meson cloud.
We find that the potential is attractive and could be strong enough to lead to the 
existence of $\Theta^+$ nuclear bound states.}
\end{abstract}

\section{Introduction}

The study of negative strangeness hypernuclei, $\Lambda$, $\Sigma$ and $\Xi$ 
has been a fruitful area of research.
It has brought valuable information on the hyperon-$N$ interaction, the hyperon-$ N \rightarrow
NN$ weak transitions, and many other interesting topics
\cite{Dover:sv,Oset:1989ey,Alberico:2001jb,Gal:si,Hashimoto:1999ks,Itonaga:2002gj}.
   The discovery of an exotic baryon with positive strangeness, $\Theta^+$ 
\cite{Nakano:2003qx},  still controversial, opens the possibility of the existence of 
positive strangeness hypernuclei, which could provide information about this new baryon
 unreachable or complementary to that obtained in elementary reactions.

 This possibility was first explored 
in Ref. \cite{Miller:2004rj} where it was  suggested that
$\Theta^+$ hypernuclei, stable against strong decay, might exist. Later, in
Ref. \cite{Kim:2004fk}, the $\Theta^+$ selfenergy in the nuclei is
calculated, using as only ingredient the  $\Theta^+ K N$ coupling. The results 
showed that the interaction coming from that source is  too weak to bind the 
pentaquark in nuclei. However, in Refs. \cite{Cabrera:2004yg,cabrerapq04}
it was shown that other  selfenergy pieces related to the coupling of the $\Theta^+$ 
resonance to a  baryon  and  two mesons could lead to a
sizable attraction, enough to produce bound and narrow $\Theta^+$ states in
nuclei. Recently, Shen and Toki \cite{toki}, using  the quark mean-field model
have also found a strong attractive potential that would produce nuclear bound
states. 

In this talk, we will discuss two of the sources of the $\Theta^+$ selfenergy.
First,  we will consider the selfenergy terms related to the  $\Theta^+ K N$ 
coupling. In a second part, we will study 
the coupling of the $\Theta^+$ to two mesons 
and a nucleon by  means  a $SU(3)$ symmetric Lagrangian\cite{madrid}, constructed 
to account for the  coupling of the antidecuplet to a baryon of the nucleon octet
and two pseudoscalar mesons.  With this  Lagrangian an attractive selfenergy is 
obtained for all the members of the antidecuplet coming from the two meson 
cloud.

\section{The pentaquark selfenergy in nuclear matter}

\begin{figure}
\begin{center}
%\centerline{\epsfxsize=1.4in\epsfbox{fig1.eps}}   
 \includegraphics[height=5.0cm]{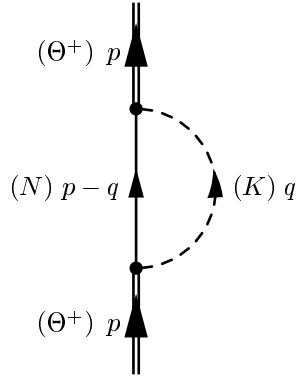}
\end{center}
\caption{$\Theta^+$  selfenergy diagram related to the $K N$ decay.}
\label{2bodyself}
\end{figure}

The $\Theta^+$  selfenergy diagram, associated to  the $KN$ decay channel,
is shown in Fig. \ref{2bodyself}.
For the $L=0$ case, the free $\Theta^+$ selfenergy from this diagram is given 
by the equation
\begin{equation}
-i\Sigma_{KN}=2\int\frac{d^4q}{(2\pi)^4} g_{K^+n}^2 
\frac{M}{E_N}
\frac{1}{p^0-q^0-E_N+i\epsilon}
\frac{1}{q^2-m_K^2+i\epsilon}\,,
\end{equation}
where $M$ is the nucleon mass, $E_N(k)=\sqrt{M^2+\vec{k}\,^2}$. The results 
for $L=1$ are obtained by the substitution 
$g_{K^+n}^2\rightarrow \bar{g}_{K^+n}^2 \vec{q}\,^2 \, .$
  
  The $\Theta^+$  selfenergy in  nuclear matter is obtained by properly modifying the
previous formula. The nucleon propagator is changed as follows
\begin{equation}
\frac{1}{p^0-q^0-E_N+i\epsilon}\rightarrow 
\frac{1-n(\vec{p}-\vec{q})}{p^0-q^0-E_N+i\epsilon}\,+\,
 \frac{n(\vec{p}-\vec{q})}{p^0-q^0-E_N-i\epsilon}\,,
\end{equation}
where $n(\vec{k})$ is the occupation number. The vacuum kaon propagator 
is replaced by the in-medium one,
\begin{equation}
\frac{1}{q^2-m_K^2+i\epsilon}\rightarrow\frac{1}{q^2-m_K^2-\Pi_K(q,\rho)}\,,
\end{equation}
where $\Pi_K(q^0,|\vec{q}|,\rho)$ is the kaon selfenergy. The $s-$wave part 
of this selfenergy is well approximated by\cite{Kaiser:1996js,Oset:2000eg} 
$
\Pi_{K}^{(s)}(\rho)=0.13\, m_K^2 \rho/\rho_0\ \,,
$
where $\rho_0$ is the normal nuclear density. The $p-$wave part is taken from 
the model of Ref. \cite{Cabrera:2002hc},
which accounts for $\Lambda h$, $\Sigma h$ and $\Sigma^*(1385) h$ excitations.

\begin{figure}
%\centerline{\epsfxsize=4.6in\epsfbox{fig2.eps}}
\begin{center}
    \includegraphics[width=0.75\textwidth]{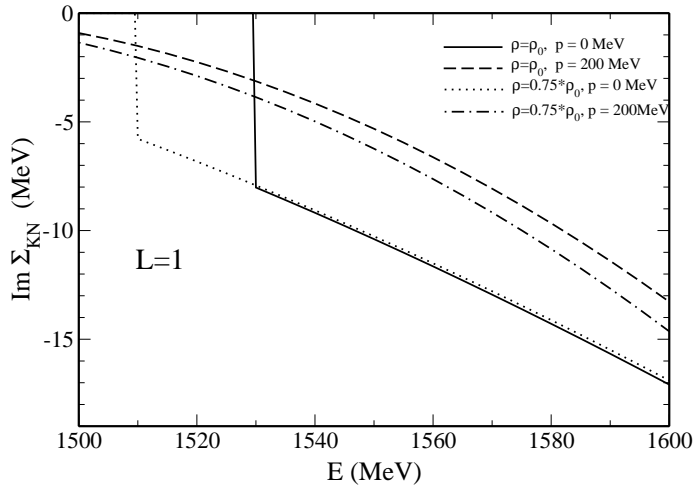}
\end{center}
\caption{Imaginary part of the $\Theta^+$ selfenergy associated to 
the $KN$ decay channel for  $L=1$.}
\label{ImL1}
\end{figure}

 We show the results for the case $L=1$ in Fig. \ref{ImL1}, where 
$\Gamma=15$ MeV has been used to obtain the coupling constants. 
The in medium selfenergy  is proportional to the vacuum width, thus the results 
should be  appropriately scaled when the width is better determined. 
 In any case, even if $\Gamma=15$ MeV in vacuum, inside the nucleus the width 
is smaller, basically because of the Pauli blocking. 
For the case $L=0$ we obtain even smaller results. A full discussion of the different
curves can be found in Ref. \cite{Cabrera:2004yg}. What should be remarked 
is that for a typical case with some 20 MeV binding and a $\Theta^+$ momentum
of 200 MeV (also typical for a nuclear state)
this width would be around 5 MeV or smaller if the $\Theta^+$ width is smaller than
15 MeV.

The  real part of the $\Theta^+$ selfenergy in the medium is obtained after 
subtraction of the vacuum one. The results coming from this source are small, 
of the order of 1 MeV  or less and not enough to bind $\Theta^+$ in nuclei.

\begin{figure}
%\centerline{\epsfxsize=3.1in\epsfbox{fig3.eps}}   
\begin{center}
\includegraphics[width=0.7\textwidth]{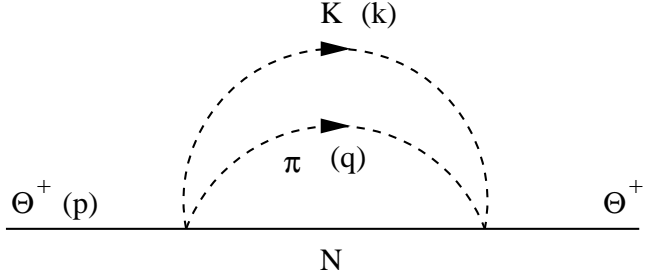}
\end{center}
\caption{\label{Threebody} Two-meson $\Theta^+$ selfenergy diagram.}
\end{figure}

Now, we will discuss 
the $\Theta^+$ selfenergy tied to the two-meson cloud.
We consider  contributions to the $\Theta^+$ selfenergy from diagrams in which
the $\Theta^+$ couples to a nucleon and two mesons, like the one in Fig.
\ref{Threebody}.
There is no direct information on these couplings since the $\Theta^+$ mass is
below the two-meson decay threshold and there are several possibilities depending
on the $\Theta^+$ quantum numbers. 

 From now on we make several assumptions.
 First, we assume the $\Theta^+$  to have spin and parity $J^P=1/2^+$ 
associated to an $SU(3)$ antidecuplet\cite{Diakonov:1997mm}. Also the
$N^*(1710)$ is supposed to couple strongly to the same antidecuplet.

 From the PDG data on $N^*(1710)$ decays we  can determine some of the couplings 
of the antidecuplet to the two-meson channels.

 In order to account for the 
$N^*(1710)$ decay into $N (\pi\pi, p-\textrm{wave},I=1)$ and $N (\pi\pi,
s-\textrm{wave},I=0)$ we use the following lagrangians
\begin{equation}
\label{L1}
{\mathcal L}_1 = i g_{\bar{10}} \epsilon^{ilm} \bar{T}_{ijk} 
\gamma^{\mu} B^j_l (V_{\mu})^k_m \, ,
\end{equation}
with
\begin{equation}
\label{veccurr}
V_{\mu} = \frac{1}{4 f^2} (\phi \partial_{\mu} \phi - \partial_{\mu} 
\phi \phi)
\, ,
\end{equation}
where $f=93$ MeV is the pion decay constant and $T_{ijl}$, $B^j_l$, $\phi ^k_m$
$SU(3)$ tensors which account for the antidecuplet states, the octet of
$\frac{1}{2}^+$ baryons and the octet of $0^-$ mesons, respectively.
 The second term is given by
\begin{equation}
\label{L2}
{\mathcal L}_2 = \frac{1}{2 f} \tilde{g}_{\bar{10}}  \epsilon^{ilm} 
\bar{T}_{ijk}
(\phi \cdot \phi)^j_l B^k_m \, ,
\end{equation}
which couples two mesons in $L=0$ to the antidecuplet and the baryon and they
are in $I=0$ for the case of two pions.  From the Lagrangian terms of Eqs.
(\ref{L1}, \ref{L2}) we obtain the transition amplitudes 
$N^*\to \pi\pi N$. Taking the central values from the PDG\cite{PDG} for the
$N^* (1710) \to N(\pi\pi, p-\textrm{wave},I=1)$  and for the 
$N^* (1710) \to N(\pi\pi,s-\textrm{wave},I=0)$, the resulting coupling 
constants are $g_{\bar{10}}=0.72$
and  $\tilde{g}_{\bar{10}}=1.9$. The uncertainties for these constants are
quite large with the current experimental information.

There are other possible pieces, with a different $SU(3)$ structure
but they are not relevant for this work. A more complete study of this
couplings can be found in Ref. \cite{madrid}.

The $\Theta^+$ selfenergy associated to the diagram of Fig. \ref{Threebody} is
given by pieces of the type
\begin{eqnarray}
\label{sigmaj}
\Sigma^j(p) &=& - \int \frac{d^4 k}{(2\pi)^4} \int \frac{d^4 q}{(2\pi)^4}
|t^j|^2 \frac{1}{k^2-m_1^2+i\epsilon} \, \frac{1}{q^2-m_2^2+i\epsilon}
\nonumber \\ & &
\frac{M}{E(\vec{k}+\vec{q})} \,
\frac{1}{p^0-k^0-q^0-E(\vec{k}+\vec{q})+i\epsilon}
\, ,
\end{eqnarray}
where $m_1$ and  $m_2$ are the masses of the mesons in the loop ($K \eta$, $K \pi$)
and  $t^j$ are the amplitudes obtained from the two lagrangian terms.

The implementation of the medium effects is done by including the medium
selfenergy of the kaon and modifying the nucleon propagator, as
shown before. On the other hand, for the pion we modify
the propagator including the $p-$wave selfenergy which is basically driven by
$ph$ and $\Delta h$ excitations and short range correlations 
\cite{Oset:1981ih,Chiang:1997di,Cabrera:2004kt}.
Once the $\Theta^+$ selfenergy at a density $\rho$ is evaluated, the optical 
potential felt by the $\Theta^+$ in the medium is obtained by subtracting the
free $\Theta^+$ selfenergy.
Obviously, the imaginary part
of the selfenergy is finite, however, the real part is divergent, even after
subtraction of the free one, and must be regularized. We do it by using a 
cutoff.

\begin{figure}
%\centerline{\epsfxsize=4.6in\epsfbox{fig4.eps}} 
\begin{center} 
 \includegraphics[width=0.75\textwidth]{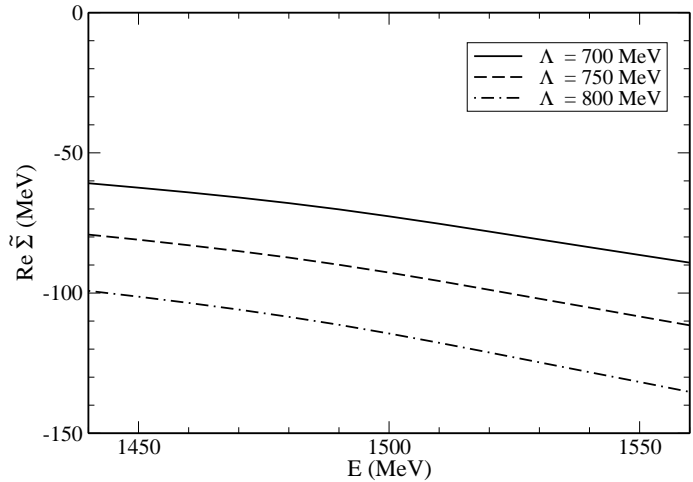}
 \end{center}
\caption{Real part of the two-meson contribution to the $\Theta^+$ selfenergy
at $\rho=\rho_0$.} 
\label{Re2meson}
\end{figure}

The results for the real part of the selfenergy are presented in Fig. 
\ref{Re2meson}. The potential for $\rho=\rho_0$ is attractive
and quite strong. 
This large effect comes mostly from the pion selfenergy, which 
also induces a very large attraction  in other cases like for the  
$\sigma$\cite{Chiang:1997di} and the $\kappa$ 
mesons\cite{Cabrera:2004kt}. The medium effects affecting 
the kaon and the nucleon are not so relevant for the final results.

Even with the quoted large uncertainties we conclude that there could be a 
sizable attraction of the order of magnitude of 50-100 MeV at normal nuclear 
density, which is more than enough to bind the $\Theta^+$ in any nucleus.

\begin{figure}
\begin{center}
%\centerline{\epsfxsize=4.6in\epsfbox{fig5.eps}}   
\begin{center}
    \includegraphics[width=0.75\textwidth]{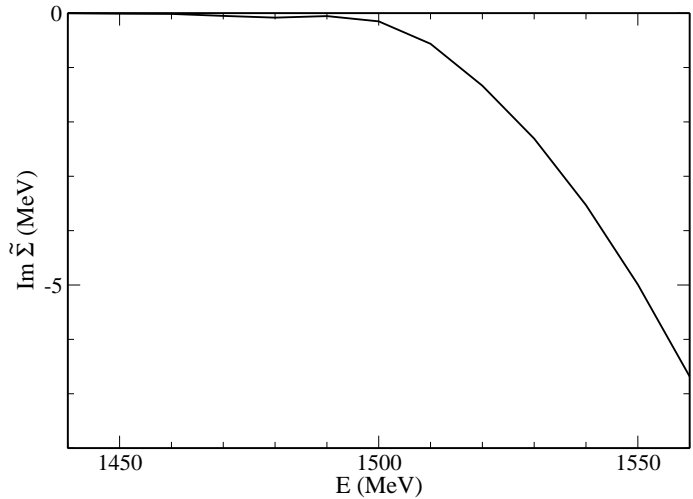}
    \end{center}
\caption{Imaginary part of the two-meson contribution to the $\Theta^+$ 
selfenergy at $\rho=\rho_0$.}
\label{Im2meson}
\end{center}
\end{figure} 
In Fig. \ref{Im2meson} we show the imaginary part of the $\Theta^+$ selfenergy 
related to the two-meson cloud. We find that $\Gamma$ would be
smaller than 5 MeV for bound states with a binding of $\sim$20 MeV and
negligible for binding energies of $\sim$40 MeV or higher.

 This, together with
the small widths associated to the $KN$ decay channel, would lead to $\Theta^+$
widths below 8 MeV, assuming a free width of 15 MeV, and much lower if the
$\Theta^+$ free width is of the order of 1 MeV, as suggested in  Refs. 
\cite{Gibbs:2004ji,Sibirtsev:2004bg} and quoted by the PDG \cite{PDG}. 
In any case, for most
nuclei, this width would be smaller than the separation of the deep 
levels\cite{Nagahiro:2004wu}.

\section{Conclusions} 

We find that although
the selfenergy of the $\Theta^+$ in  nuclear matter
associated to the $KN$ decay channels is quite weak, there is a large
attractive $\Theta^+$ potential in the nucleus associated to the two meson
cloud of the resonance. Such a strong potential would imply the existence of
$\Theta^+$ hypernuclei.

These states would decay through the $KN$ channel, but also via
 new decay channels open  in the
medium, like $\Theta^+ N\rightarrow NNK$. The full width 
taking into account the new medium channels and 
the $KN$ decay, is quite small compared to the
separation of the energy levels of the $\Theta^+$ pentaquark 
in light and intermediate nuclei. 

Our results are based on some assumptions about the $\Theta^+$ 
quantum numbers, the composition of the antidecuplet and 
the possible couplings to the two mesons and a octet baryon.
Additionally, there are large experimental uncertainties
in the data used to fix the size of the couplings.
Nonetheless, the potential is so large that even allowing for 
a wide margin of uncertainty, we think the possibility of
existence of bound states is well founded.

\section*{Acknowledgments}
This work is partly supported by DGICYT contract number BFM2003-00856,
the E.U. EURIDICE network contract no. HPRN-CT-2002-00311 and by the
Research Cooperation Program of the Japan Society for the Promotion of Science
(JSPS) and the spanish Consejo Superior de Investigaciones Cientificas (CSIC).
D.~C. acknowledges financial support from MCYT and Q.~B.~Li acknowledges 
support from the Ministerio de Educaci\'on y Ciencia in the program of Doctores
y Tecn\'ologos Extranjeros.

\end{document}